\documentclass[conference]{IEEEtran}

\usepackage[utf8]{inputenc}
\usepackage[T1]{fontenc}
\usepackage{newtxtext}
\usepackage[scaled=0.85]{inconsolata}
\usepackage{microtype}
\usepackage{graphicx}
\graphicspath{{figures/}}
\newcommand{\resStrongRestPct}{86}
\newcommand{\resStrongGraphPct}{88}

\newcommand{\pctStrongRestXsTriage}{79}

\newcommand{\tokStrongRestXs}{28}

\newcommand{\tokStrongGraphXs}{36}

\newcommand{\pctWeakRestXsRca}{29}

\newcommand{\tokWeakRestXs}{29}

\newcommand{\pctWeakGraphXsRca}{77}

\newcommand{\tokWeakGraphXs}{29}
\newcommand{\pctStrongRestLargeImpact}{52}

\newcommand{\pctStrongRestLargeTriage}{100}
\newcommand{\pctStrongRestLargeAll}{78}
\newcommand{\tokStrongRestLarge}{39}

\newcommand{\pctStrongGraphLargeAll}{100}
\newcommand{\tokStrongGraphLarge}{35}

\newcommand{\pctWeakRestLargeRca}{23}

\newcommand{\tokWeakRestLarge}{61}

\newcommand{\pctWeakGraphLargeRca}{67}

\newcommand{\tokWeakGraphLarge}{44}
\newcommand{\scaleFilesRatio}{15}
\newcommand{\tasksSmaller}{72}

\newcommand{\tasksSmallerAnyOf}{7}
\newcommand{\tasksSmallerMulti}{20}
\newcommand{\tasksSmallerAbstain}{1}
\newcommand{\tasksLarger}{72}

\newcommand{\tasksLargerAnyOf}{0}
\newcommand{\tasksLargerMulti}{17}
\newcommand{\tasksLargerAbstain}{0}

\newcommand{\precFlaggedMax}{20}
\newcommand{\precOthersMin}{61}
\newcommand{\precFlagPct}{50}
\newcommand{\trialMaxDiff}{7}

\newcommand{\strongRestCalls}{18.7}
\newcommand{\strongRestResultK}{18.8}

\newcommand{\strongGraphCalls}{8.5}
\newcommand{\strongGraphResultK}{25.3}

\newcommand{\traceEvidenceStrongRest}{99}
\newcommand{\traceDecisionStrongRest}{20}
\newcommand{\traceEvidenceStrongGraph}{100}

\newcommand{\traceEvidenceWeakRest}{31}

\newcommand{\traceEvidenceWeakGraph}{44}

\newcommand{\traceLongestRetrieval}{31}
\newcommand{\probeRestIdOk}{100}\newcommand{\probeRestIdN}{100}
\newcommand{\probeGraphLabelOk}{0}\newcommand{\probeGraphLabelN}{80}
\newcommand{\probeGraphIriOk}{80}\newcommand{\probeGraphIriN}{80}

\newcommand{\corpusSmallerServedfiles}{728}
\newcommand{\corpusSmallerCustomers}{5}
\newcommand{\corpusSmallerServices}{5}

\newcommand{\corpusSmallerResources}{180}
\newcommand{\corpusSmallerAlarms}{158}
\newcommand{\corpusLargerServedfiles}{10{,}706}
\newcommand{\corpusLargerCustomers}{21}
\newcommand{\corpusLargerServices}{22}

\newcommand{\corpusLargerResources}{6{,}480}
\newcommand{\corpusLargerAlarms}{3{,}006}
\newcommand{\corpusSmallerTriples}{14{,}154}
\newcommand{\corpusSmallerMegabytes}{8}
\newcommand{\corpusLargerTriples}{151{,}630}
\newcommand{\corpusLargerMegabytes}{321}

\newcommand{\entailSmallerTriples}{3{,}524}
\newcommand{\entailSmallerPct}{25}
\newcommand{\entailSmallerQueries}{2{,}442}
\newcommand{\entailSmallerEmptyToNon}{2}
\newcommand{\entailSmallerChanged}{840}
\newcommand{\entailSmallerAgree}{1{,}883}
\newcommand{\entailSmallerCompared}{1{,}899}
\newcommand{\entailLargerTriples}{45{,}857}
\newcommand{\entailLargerPct}{30}
\newcommand{\entailLargerQueries}{58}
\newcommand{\entailLargerEmptyToNon}{0}
\newcommand{\entailLargerChanged}{12}

\newcommand{\tboxTriples}{17{,}860}
\newcommand{\overlayTriples}{70}
\newcommand{\ctsTboxTriples}{11{,}958}
\newcommand{\nrmTboxTriples}{28{,}588}

\newcommand{\genSmallerCells}{96}

\newcommand{\genSmallerFiles}{826}
\newcommand{\genLargerCells}{2{,}720}

\newcommand{\genLargerFiles}{11{,}124}
\newcommand{\genMCells}{7{,}200}
\newcommand{\genMSeconds}{1{,}854}

\newcommand{\genMRatio}{2.6}

\newcommand{\joinsDeclared}{18}
\newcommand{\joinsTotal}{12}

\newcommand{\joinsHopsDeclared}{7}
\newcommand{\joinsHopsTotal}{34}

\usepackage{booktabs}
\usepackage{tabularx}
\usepackage{hyperref}
\usepackage{amsmath,amssymb,amsthm}
\usepackage{newtxmath}

\usepackage{algorithm}
\usepackage{algorithmic}
\usepackage{multirow}
\usepackage{listings}
\usepackage{xcolor}
\usepackage{balance}
\usepackage[acronym,nonumberlist,nogroupskip]{glossaries}

\newacronym{sid}{SID}{Shared Information and Data}
\newacronym{tmf}{TMF}{TM Forum}
\newacronym{snm}{SNM}{semantic network management}
\newacronym{tio}{TIO}{TM Forum Intent Ontology}
\newacronym{oda}{ODA}{Open Digital Architecture}
\newacronym{kg}{KG}{Knowledge Graph}
\newacronym{cfs}{CFS}{Customer-Facing Service}
\newacronym{rfs}{RFS}{Resource-Facing Service}
\newacronym{nrm}{NRM}{Network Resource Model}
\newacronym{asgi}{ASGI}{Asynchronous Server Gateway Interface}
\newacronym{shacl}{SHACL}{Shapes Constraint Language}

\newacronym{cts}{CTS}{Common Topology Service}
\newacronym{rdf}{RDF}{Resource Description Framework}
\newacronym{rdfs}{RDFS}{RDF Schema}
\newacronym[first={REST},firstplural={REST}]{rest}{REST}{Representational State Transfer}
\newacronym{iri}{IRI}{Internationalized Resource Identifier}
\newacronym{llm}{LLM}{Large Language Model}
\newacronym{ci}{CI}{confidence interval}
\newacronym[first={API},firstplural={APIs}]{api}{API}{Application Programming Interface}

\makeglossaries

\makeatletter
\renewcommand*{\glsdohyperlink}[2]{{\hypersetup{pdfborder={0 0 0}}\hyperlink{#1}{#2}}}
\let\@glslink\glsdohyperlink
\makeatother

\usepackage{listings}
\usepackage{listings-turtle-paper}
\usepackage{listings-sparql-paper}

\lstdefinelanguage{JSONPaper}{
  morestring=[b]",
  stringstyle=\color{tp-string},
  morecomment=[l]{//},
  commentstyle=\color{tp-comment},
  literate={\{}{{{\color{tp-punct}\{}}}1 {\}}{{{\color{tp-punct}\}}}}1
           {[}{{{\color{tp-punct}[}}}1 {]}{{{\color{tp-punct}]}}}1
           {:}{{{\color{tp-punct}:}}}1 {,}{{{\color{tp-punct},}}}1,
}
\lstdefinelanguage{ShellPaper}{
  morecomment=[l]\#,
  commentstyle=\itshape\color{tp-comment},
  morestring=[b]",
  stringstyle=\color{tp-string},
  keywords={curl,uv,run,make,cd,export},
  keywordstyle=\bfseries\color{tp-keyword},
}

\lstdefinestyle{paperlisting}{
  basicstyle=\ttfamily\footnotesize,
  breaklines=true,
  columns=fullflexible,
  numbers=left,
  numberstyle=\tiny\color{tp-comment},
  numbersep=5pt,
  xleftmargin=1.6em,
  tabsize=2,
  showstringspaces=false,
  upquote=true,
  frame=single,
  framesep=3pt,
  rulecolor=\color{tp-punct},
  aboveskip=0.6em,
  belowskip=0.4em,
}
\definecolor{layer-api}{RGB}{255,255,255}
\definecolor{layer-sem}{RGB}{255,255,255}
\definecolor{layer-kg}{RGB}{255,255,255}

\begin{document}

\title{Semantic Network Management: What Standardized APIs Leave Unsaid, and What It Costs AI Agents}

\author{
\IEEEauthorblockN{Jean Martins\IEEEauthorrefmark{1}, Amardeep Kumar A\IEEEauthorrefmark{2},
Marin Orlic\IEEEauthorrefmark{1}, Dinand Roeland\IEEEauthorrefmark{1}}
\IEEEauthorblockA{\IEEEauthorrefmark{1}Ericsson Research \quad \IEEEauthorrefmark{2}Ericsson BCSS\\
\{jean.martins, amardeep.a.kumar, marin.orlic, dinand.roeland\}@ericsson.com}
}

\maketitle
\hypersetup{pdftitle={Semantic Network Management: What Standardized APIs Leave Unsaid, and What It Costs AI Agents}}

\begin{abstract}
Telecommunications management relies on standardized interfaces spanning business, operations, and network domains. However, different standards development organizations often publish information models whose concepts, relationships, and identifiers are not fully aligned: an alarm and an inventory record may name the same cell differently, and no standard states that they are the same.
Traditionally, such mismatches have been resolved through integration logic hidden behind the interfaces. We argue that this creates a semantic gap that obscures domain knowledge from AI agents and should instead be addressed by a shared semantic layer over the interfaces. In this work, we investigate how to mitigate such gaps by semantically lifting the information exposed by existing interfaces.
To evaluate this premise, we generate the payloads of those standardized interfaces for a synthetic 5G network. The resulting dataset captures a consistent network state over which root-cause analysis, customer-impact assessment, and customer-triage tasks can be defined.
We then perform a semantic lifting process that transforms schemas and payloads into an RDF knowledge graph, explicitly encoding equivalences across identifiers, entities, and interfaces.
We evaluate the same tasks using an AI agent with data access MCP tools, under a stronger and a weaker model (GPT-5.6-sol and GPT-4.1), over standard REST \glspl{api}, which leave cross-system semantics unsaid, and SPARQL on the lifted graph, which states them, across two corpora differing in size by a factor of \scaleFilesRatio.
With the same task guidance for both interfaces, the graph halves the data calls of the stronger model (from \strongRestCalls\ to \strongGraphCalls\ per task) and, on the larger corpus, lets it solve every task, where REST solves \pctStrongRestLargeAll\%; the gap lies in cross-domain customer impact. The weaker model gains mainly on root-cause analysis (from \pctWeakRestXsRca\% to \pctWeakGraphXsRca\%) but fails customer impact over both interfaces. As the corpus grows, the stronger model's cost per solved task rises over REST and stays flat over the graph.
Our results indicate that a semantic layer that states cross-domain semantics can improve the robustness and efficiency of AI agents, suggesting that standardized semantic modeling and semantic network management may pave the way for the agentic AI era.
\end{abstract}


\section{Introduction}
\label{sec:introduction}

Consider an alarm on a radio cell. To determine which customer-facing service it affects, and whether that service’s latency objective is still being met, a management application must trace that cell across four interfaces. The network manager that raised the alarm identifies it by a 3GPP distinguished name; the topology interface represents the same cell under a different identifier; the service inventory relates it to a resource-facing service and the customer-facing service above it; and the assurance interface exposes the metric against which the objective is judged. Each interface assigns its own types and identifiers, but the published schemas do not state that these descriptions refer to the same entity. In practice, the match is made through integration code, written for each consumer and revised whenever either side changes. The burden grows in multi-vendor networks, where each vendor names its own parameters and several vendors may serve the same customer.

This arrangement has worked in the past because consumers, meaning the applications and systems that call the interfaces, were integrated in advance. An engineer read the documentation, determined which identifiers correspond across interfaces, a correspondence we call a \emph{join}, and encoded that knowledge. AI agents driven by \glspl{llm} can in principle do a similar job, but doing so makes each agent responsible for inferring correspondences that are the same for every consumer of the network. We argue that such joins, a form of cross-domain semantics, should be managed as network knowledge, i.e., if agents are to reason directly over standardized interfaces, the standards, or a shared semantic layer above them, should expose them.

We use the term \emph{semantic network management}~\cite{matheus2019snm} referring to a paradigm in which management data and knowledge are machine-interpretable at the point of access. The term is rooted in the Semantic Web, which made the meaning of data readable by machines through ontologies and linked identifiers~\cite{berners-lee2006linkeddata}, and has gained new traction because AI agents now use graphs in a similar way, to retrieve and assemble the context they reason over~\cite{edge2024graphrag,singh2025agenticrag,mei2025context}. The standards are moving the same way, where the \gls{tio}~\cite{tio36} expresses intent in \gls{rdf}, and TM Forum's TR326~\cite{tr326} proposes operationalizing ontologies over its Open APIs for AI-native operations.

Telecom is unusually well placed for this. In many domains, knowledge graphs for AI agents must be built by extracting entities and relations from unstructured text, because no formal description of the domain exists; that extraction is itself a source of error. Telecom, by contrast, is among the most heavily standardized domains and much of its structure is already formalized in APIs and information models. The challenge is therefore not to acquire that structure from prose, but rather to make it explicit, mechanically, auditably, and repeatably, across versions of a standard and domains.

What stands in the way of \gls{snm} is one gap with two sides. On the specification side, standards bodies such as \gls{tmf} and 3GPP publish the information models as schemas rather than as machine-interpretable semantics. On the implementation side, much of the meaning the standards do define is written in prose documentation, which no API payload describes, and the integration code that supplies it is usually inside the products of telecom vendors and operators, out of reach of researchers. As a result, most related literature on telecom knowledge graphs and semantic models~\cite{krinkin2020kg,noria2024,kulikov2021ontology,gspec2025} stays at some distance from what could actually be tried on today's systems, and none measures how much a consumer can recover from the published schemas alone, or what must be added.

This paper addresses both sides of that gap in cross-domain 5G management. We employ a data generator that produces the payloads of the standardized interfaces together with the joins between them as ground truth, and a semantic lifting process that extracts the semantics of the published schemas and asserts the joins in an RDF knowledge graph (henceforth \textit{semantic layer}), through authored rules, e.g., a distinguished name identifies the same object in every system. Such infrastructure is then used to measure the pros and cons of the semantic layer from different angles.

The paper makes three contributions:

\begin{enumerate}
\item A semantic lifting process.
It derives ontologies from the \gls{tmf} JSON Schema corpus\footnote{\url{https://github.com/tmforum-apis/Open_Api_And_Data_Model}} (\tboxTriples\ triples) and from the schema of a real telecom vendor's internal service topology and resource inventory API.
It lifts API payloads into a graph through JSON-LD without changing the APIs, using type-scoped contexts\footnote{\url{https://www.w3.org/TR/json-ld11/\#scoped-contexts}}: each class gets its own mapping from JSON keys to \glspl{iri}. Authored rules assert the joins in the graph as \texttt{owl:sameAs} links, so that a single query can follow the chain from the alarm to the customer.

\item A measurement of how far today's standards take a consumer on their own. On a synthetic 5G network modelled on real vendor products and standards, \joinsTotal\ of the declared joins are unstated: both identifiers appear in the payloads, but no payload says that they correspond. We test whether a consumer can derive them mechanically from the published schemas. None can, since no schema declares that the identifiers link the systems, so this knowledge lives outside the schemas, and a consumer needs either rules authored from domain knowledge to state it or its own inference at run time.

\item A comparison of two interfaces over the same generated data at two sizes, served by REST APIs vs a centralized RDF graph served via SPARQL. Those are put to a stronger and a weaker model (GPT-5.6-sol and GPT-4.1), with a general-purpose agent as a measuring instrument to analyse accuracy, cost and the traces of how each interface is used.

\end{enumerate}

Section~\ref{sec:background} reviews the background and related work, Section~\ref{sec:snm} the path from Linked Data to semantic network management, Section~\ref{sec:architecture} the lifting, Section~\ref{sec:method} the corpus and method, Section~\ref{sec:results} the results, and Section~\ref{sec:conclusion} concludes.

\section{Background and Related Work}
\label{sec:background}

\subsection{Standardized Network and Service Management Interfaces}

The \gls{tmf} \gls{sid} information model defines the main entities of telecom operations, including customers, products, services, and resources, together with their relationships. In \gls{oda}, these models are exposed through REST \glspl{api}, the TM Forum Open APIs, with each software component publishing its capabilities as OpenAPI-described interfaces such as TMF629 (Customer Management), TMF637 (Product Inventory), TMF638 (Service Inventory), TMF639 (Resource Inventory), TMF642 (Alarm Management), and TMF641 (Service Ordering).

In practice, applications read inventories through filtered, paged GET requests, and request changes, such as creating or modifying a service, by submitting an order. Finding the customers affected by an alarm, for example, takes a chain of reads across the alarm (TMF642), the referenced resource (TMF639), the dependent services (TMF638), and the products and customers (TMF637, TMF629), matching string identifiers in integration logic external to the APIs. The APIs thus give \emph{syntactic} interoperability, common JSON resource structures, but no machine-readable semantics, i.e., nothing formally declares that these identifiers denote the same entity, nor that the service's latency objective is the quantity the assurance system measures for it.

Similarly, 3GPP covers the network domain. Its \gls{nrm}~\cite{3gpp28541} defines radio and core managed objects as information object classes, with YANG and OpenAPI solution sets, and identifies each object instance by a distinguished name. Vendors, in turn, add interfaces of their own, such as the service topology and resource inventory API used in our corpus.

An application that spans customer, service, and network domains must therefore combine APIs from several standardization lineages, each with its own modeling conventions, and rely on integration code to relate them.

\subsection{From Intents and Agents to Semantics}
\label{sec:intents}

Intent-based management lets an operator state the outcome it wants, such as a service level for a customer or an area, and leaves the network to decide how to realize and keep it~\cite{leivadeas2023survey,ericsson-evolution}. Intent standards differ primarily in representation. TM Forum's approach is RDF-native: TMF921~\cite{tmf921} defines the intent lifecycle, while the \gls{tio}~\cite{tio36} provides an \gls{rdfs} vocabulary in which intents can be expressed and validated with SHACL shapes~\cite{martins2026tioshacl}. By contrast, 3GPP's intent-driven management~\cite{3gpp28312} models intents as information object classes within the \gls{nrm} with OpenAPI solution sets.

In both families, an intent refers to domain objects defined elsewhere, such as services, cells or measurements~\cite{tio36}. Even the RDF-native \gls{tio} therefore lacks a machine-interpretable representation of the objects its intents target. An intent expressed in RDF then names objects that the same standards leave outside it, which limits what can be checked or inferred about the intent, such as whether the object it targets exists, or is the object another system reports.

In recent years intents have moved from formal RDF models such as the \gls{tio}~\cite{tio36,martins2026tioshacl} and the structured information object classes of 3GPP~\cite{3gpp28312} to natural-language intents, which no standard defines and which \glspl{llm} must translate and ground in the network~\cite{swift2024,martins2026intent}. As the closed loops that realize intents~\cite{gomes2021intent,mehmood2025intent,ericsson-evolution} are replaced by \gls{llm}-driven ones, curated semantics gives their inferences an anchor in the objects of the existing network, as formal vocabulary where the model is formal, and as context and knowledge where an \gls{llm} or an agent interprets the request~\cite{karapantelakis2024genai}.

The importance of semantics grows with agentic AI and higher levels of autonomy, because an agent that consumes standard interfaces has no human integrator to reconstruct them for it, and depends on what the standards state, or on inferring it at every run~\cite{tranoris2026agentic,tacheny2026diagnostic}. General-purpose agent designs have turned to knowledge graphs and curated context for the same reason, using them to structure what a model retrieves and reasons over~\cite{pan2024unifying,edge2024graphrag}, to decide at each step what to fetch~\cite{singh2025agenticrag} and to assemble and govern what it sees~\cite{mei2025context,yuan2025kgrag}. Agents therefore put direct pressure on context integration and retrieval efficiency, which strengthens the case for richer semantics in management standards.

\subsection{Related Work}
\label{sec:related-kg}

\emph{Standards and API semantics.} Interoperability across vendors comes from the standards and not from each deployment's own model, so deriving semantics from them has a long precedent. Schema translations map ontologies to REST APIs, or schemas such as \gls{sid}, OpenAPI and YANG to ontologies~\cite{garijo2020oba,espinoza2020owl2openapi,muhamad2022openapi,ieva2022yang2owl,liu2006sid}, and ontology mapping was proposed early to reconcile network-management models~\cite{wong2005mapping}, but these carry types and properties, not which records in two systems are the same entity. Mapping and linked-data layers~\cite{rml2014,lanthaler2012jsonld,polli-restapi-ld,stolk2022tennet,ngsild2024,smartdatamodels} can express such joins, only through mappings or contexts written per source, whereas we derive classes and properties from the published schemas and author only the identity rules. Standards bodies are moving the same way, with IETF drafts on knowledge graphs built from network data and YANG models~\cite{nmopkgconstruct,nmopkgnetops,noriadraft}, and TM Forum's TR326~\cite{tr326}, which proposes a semantic layer over its Open APIs but leaves identifier reconciliation to unspecified services. None of these measures its value to a consumer.

\emph{Telecom knowledge graphs and agents.} Graphs for network monitoring, incident management, autonomic SDN and agentic 5G, and knowledge-graph retrieval for telecom tasks~\cite{krinkin2020kg,noria2024,zhou2021seanet,gspec2025,yuan2025kgrag}, build their ontologies for the purpose instead of deriving them from the standards, so they need re-integration to work across vendors, and none aligns \gls{tio}, \gls{sid} and 3GPP \gls{nrm} in one pipeline. General agent designs that retrieve over graphs and engineer their context~\cite{edge2024graphrag,singh2025agenticrag,mei2025context} must first construct that knowledge from text~\cite{zhu2023llmkgc,pan2024unifying}, and in telecom, symbolic techniques avoid that at a high curation cost~\cite{karapantelakis2024genai}.

\emph{What is not covered.} G-SPEC~\cite{gspec2025} and Yuan et al.~\cite{yuan2025kgrag} measure through a consumer, but neither lifts unmodified \glspl{api} to RDF from the standard schemas. We do, measuring through a probe what lifting buys and what it does not.

\section{Toward Semantic Network Management}
\label{sec:snm}

Berners-Lee's five-star scheme for Linked Data~\cite{berners-lee2006linkeddata} rates published data against five criteria, namely (1) availability on the web under an open licence, (2) structured, machine-readable data, (3) a non-proprietary format, (4) stable URIs that make entities cross-referenceable, and (5) links to other data that give context. TMF, 3GPP and vendor payloads satisfy (2) and (3), being structured data in JSON and XML, but not (4), since their classes and properties are not named by URIs, nor (5), since identifiers do not link to data in other systems, meaning the network manager, the topology, the service inventory or assurance, each with its own interface and information model. The first criterion concerns licensing and we leave it aside.

\begin{figure*}[t]
\centering
\includegraphics[width=\textwidth]{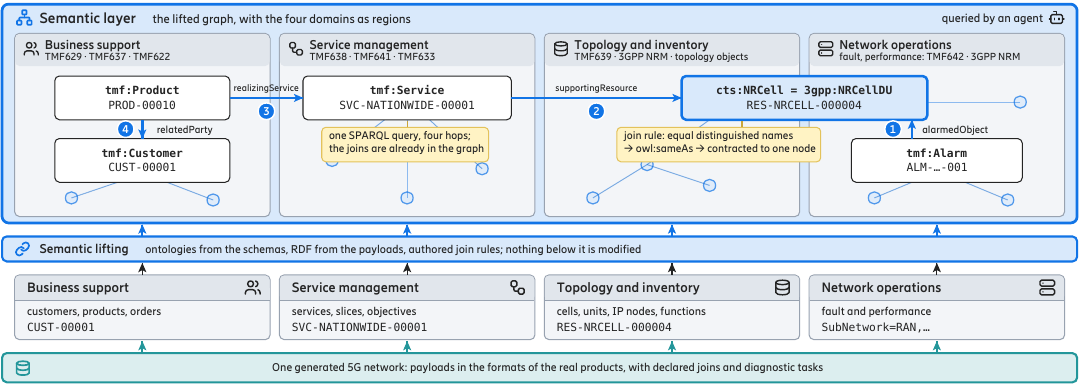}
\caption{The semantic layer in context. Bottom: one generated 5G network (teal) stands in for the managed systems, each of which publishes its own interface and names things its own way (grey). Middle: semantic lifting reads those interfaces unmodified, derives ontologies from their schemas and lifts their payloads into a graph; nothing below it is changed. Top: the graph, with the same four domains as regions. A single SPARQL query answers the question from Section~\ref{sec:introduction}, ``Which customers does this alarm affect?'', by traversing four hops (1--4), each crossing a domain boundary; the authored join rule has made the alarm's cell and the topology object a single node. Colour marks what this work adds, the semantic layer in blue and the generated dataset in teal; grey is standard.}
\label{fig:overview}
\end{figure*}

Criteria (4) and (5) are the ones management payloads miss, and we turn them into a three-level \emph{semantic maturity} ladder, whose levels are what standards would publish as semantic modeling is gradually adopted. At Level~1, corresponding to (4), a specification names its classes and properties with stable, universally unique identifiers that third parties can cite. At Level~2, corresponding to (5) within one specification, its references are typed, so that they can be traversed without prior knowledge of which endpoint stores the target record. At Level~3, corresponding to (5) across specifications, the correspondences between specifications are declared: two identifiers, or two names for a measured quantity, are stated to denote the same referent, either as identity~\cite{halpin2010sameas} or through an alignment~\cite{euzenat2013ontology}.

\subsection{Class Identification and Typed References}
\label{sec:division-mechanical}

\textit{Level~1}. Class and property identity is a naming problem, known in Linked Data as minting identifiers~\cite{hogan2021kg}. Information models define names that are unique within their own scope, but not across specifications. A specification would solve it once, by prefixing every class and property name with a stable, versioned string of its own, which gives each an \gls{iri} that third parties can cite.

\textit{Level~2}. Traversable references are a typing problem, the core of translating a schema into an ontology~\cite{hogan2021kg}. A \texttt{\$ref} in JSON Schema is already a cross-reference; what it lacks is an explicit range, the class of entity it points to. A specification would declare it, turning a string-valued field into a typed reference, an object property whose value is an entity of a known type, which a query can follow in one \emph{hop} from one node of the graph to the next.

\subsection{Opaque Identifiers and Cross-Specification Typed References}
\label{sec:division-boundary}

\begin{figure*}[t]
\centering
\includegraphics[width=\textwidth]{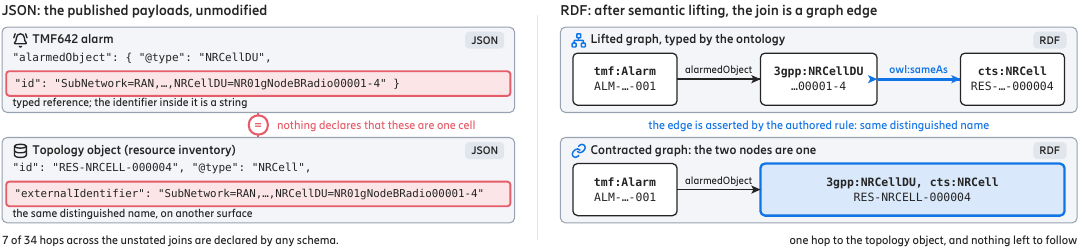}
\caption{Where the published schemas stop, and what semantic lifting adds. Left: in the published JSON, the alarm's object is a typed reference (class \texttt{NRCellDU}) whose identifier is a string, and the topology object, of a different class (\texttt{NRCell}), carries the same distinguished name, but no schema declares that both denote the same cell. Right: in the lifted RDF, nodes are typed by ontology classes (prefixes abbreviated), and semantic lifting applies the authored rule, equal distinguished names, as an \texttt{owl:sameAs} edge; under contraction, the two nodes become one node typed by both classes.}
\label{fig:join-lift}
\end{figure*}

\textit{Level~3}. Requires semantics that cross specification boundaries, which no single specification can declare. A published schema at Level~2 can take a consumer along typed references to a typed object within a single specification, but not across specifications. An alarm references its alarmed cell by its 3GPP distinguished name, a string. The service topology model of another system carries the same string, as a mapping to its own cell. The two systems agree on the identifier by construction, yet their data models each redefine a Cell class, with no equivalence and no class hierarchy or taxonomy shared between them. References across specifications are therefore shallow: they carry no semantic information that a query language, such as SPARQL, can explore (see Figure~\ref{fig:overview} and Figure~\ref{fig:join-lift}).

The gap can be closed along different paths. The simplest aligns what already exists, declaring that two records denote one entity, known as entity resolution~\cite{hogan2021kg}, or that two classes or quantities mean the same, known as ontology alignment~\cite{euzenat2013ontology}. The other path is more idealistic: it defines or reuses higher-level telecom concepts that multi-vendor specifications from different standards bodies would share, so that, for example, any Cell class would specialize a common root class.

\subsection{What Can Be Derived Today}
\label{sec:derived-today}

Until standards publish these levels, a practical route is a process that adds the missing semantics to what is published today. Levels~1 and~2 need nothing beyond the schemas: the identifier of a class or property is a deterministic function of the published specification, and the range of a reference can be recovered from its target, so nothing is authored. Level~3 cannot be derived, because no schema states the correspondences, but the simpler path above, aligning what already exists, can be taken by authoring them from domain knowledge, as rules; the shared concepts of the idealistic path remain for the standards.

\section{Semantic Lifting Architecture}
\label{sec:architecture}

Section~\ref{sec:snm} described the problem at each level, class identities and typed references within a specification (Levels~1 and~2) and opaque identifiers across specifications (Level~3), and semantic lifting solves them without changing upstream systems. Each API publishes a schema and serves payloads, and lifting uses both. The subsections below describe how this is achieved, first for schemas and then for payloads.

\subsection{Lifting Schemas: Class Identities and Typed References}
\label{sec:pipeline}

JSON-LD~1.1~\cite{jsonld11} is a concrete RDF syntax in which a context maps the keys of a JSON document to IRIs, so a document read through a context becomes a set of triples. The same process applies for TMF SID schemas and a telecom vendor's internal service topology and resource inventory API. The 3GPP resource model enters as an RDF vocabulary derived by a companion tool. Lifting runs in two stages.

\paragraph{$\textsc{Adapt}$} Annotates each schema with per-property IRIs \texttt{ontology/<domain>/<class>/<property>}, where the domain names the specification (Level~1), and resolves two \gls{tmf} structural patterns: inheritance (\texttt{allOf} with \texttt{\$ref}) becomes \texttt{rdfs:subClassOf}, and cross-references become typed references, object properties whose range is the referenced class (Level~2). %

\paragraph{$\textsc{Extract}$} Emits, for each property, its type, its domain class and a range inferred from the JSON Schema type (a string becomes \texttt{xsd:string}, an object a class IRI). The ontology is then the union of that output with an authored overlay of \overlayTriples\ integration triples (Level~3), which state what the schemas cannot. The overlay states, for example, that \texttt{dLMaxThptPerUE}, a TMF service characteristic, is the optional parameter with which a slice order sets the maximum downlink speed per user, and that it is a more specific form of the 3GPP property \texttt{dLThptPerUE}, stated as a sub-property link and not as an equivalence, so that the graph does not overstate what the data warrants.

\subsection{Lifting Payloads: Linking Identifiers Across Specifications}
\label{sec:payloads}

Payloads are lifted by a preprocessing pass that produces JSON-LD, which RDFLib then parses. This is knowledge graph construction from semi-structured data~\cite{hogan2021kg}. For each object, the pass resolves every key to the property IRI that the schema lifting assigned to it, and names the object by the same kind of convention, \texttt{instances/<class>/<id>}, from its class and the \texttt{id} it carries. It also resolves each \gls{tmf} reference object (a \texttt{*Ref}, such as \texttt{ResourceRef}) to the entity it points at, taking the type from \texttt{@referredType} or from \texttt{@type} without the \texttt{Ref} suffix. A reference and its target therefore share one IRI, and the reference becomes a link between two nodes.

Cross-specification references cannot be extracted from the payloads, so the lifting applies authored rules, each stating that two objects carrying the same distinguished name denote the same entity, and asserts an \texttt{owl:sameAs} link between them, the identity relation, since equal distinguished names are evidence of identity (Figure~\ref{fig:join-lift}). This does what an operator's integration code does when it matches identifiers, but as explicit semantics that any consumer can query in the knowledge graph. To simplify SPARQL querying, we then contract the assertion: the two nodes become one, typed by both classes, both identifiers are kept as \texttt{rdfs:label}, so that a query reaches the cell in a single node and does not have to follow \texttt{owl:sameAs} links. This is the standard canonicalisation of \texttt{owl:sameAs}~\cite{hogan2021kg}.

\section{Corpus and Evaluation Methodology}
\label{sec:method}

The evaluation asks how far the published schemas take a consumer toward semantic network management, and what a semantic layer over them is worth to an AI agent. Answering both requires a set of tasks and ground truths that real operational data rarely provides. We therefore use a proprietary data generator that emits payloads in the formats of the standards and products, declares the joins between them, and poses diagnostic questions with declared answers, so that agents' responses can be scored against them.

\subsection{Generating the Corpus and Declaring Joins}
\label{sec:method-corpus}

The data generator emits six views of a single 5G network, each in the format of a corresponding real system. Four follow the standards: the \gls{tmf} Open APIs (TMF629, TMF633, TMF637, TMF638, TMF639, TMF641, and TMF642) over the \gls{sid} model; 3GPP bulk configuration-management files, in the XML format of TS~32.615 carrying the \gls{nrm} of TS~28.541; and RAN and core performance-management result files, in the measurement-collection XML of TS~32.435. Two are the vendor's: its service topology and resource inventory model, adapted from the 3GPP \gls{nrm} and published as JSON under a JSON Schema, and its assurance definitions, thresholds and contexts, for which no schema is published. The corpus is structurally realistic where formats, identifier grammars and joins follow the products but values are synthetic but consistent.

A \emph{join} is a declared correspondence between a field in one payload and a field in another, with a match rule, a cardinality, and an authority, the specification or product behaviour that justifies it, in the spirit of the join conditions of RML~\cite{rml2014}. For example, one join equates the \texttt{externalIdentifier} of a TMF639 resource with the distinguished name carried by the topology model for the same object, by string equality, one-to-one. Some joins are \emph{unstated}: both fields appear in the payloads, but no payload says that they correspond, so a consumer must supply the correspondence from outside the payloads. The generator declares the joins of each corpus by construction and lists them (Table~\ref{tab:task-mix}), so its list is the ground truth for the whole evaluation: it says which joins exist and which are unstated. %

\subsection{Generating Diagnostic Tasks with Declaring Ground Truth}
\label{sec:method-answers}

The generator injects incidents into the network. An incident is a fault in a managed object that raises alarms (TMF642) on the cells it serves, as in the opening example, each alarm naming its cell by its 3GPP class and distinguished name, and degrades the services (TMF638) and customers (TMF629) that depend on those cells. For each incident, the generator emits one diagnostic task, with a question and its declared answer, which is what a correct correlation of the alarms would conclude, not what any particular implementation did conclude. In some cases the cause is silent: the failing node raises no alarm, while the cells it serves do, so the only route to the cause passes through the service-topology model, which is what the joins in Section~\ref{sec:method-corpus} make answerable.

There are three task types, with 24 tasks each. \emph{Root-cause analysis} asks what caused a set of cell alarms (TMF642) and on which managed object, named by its 3GPP distinguished name. \emph{Customer impact} asks which customers (TMF629) and services (TMF638) the alarms affect, the question of the opening example. \emph{Customer triage} asks what caused a given customer's degraded service.

For each type, the generator also declares a floor, a lower bound to compare agents' accuracy. A result at or below that floor is a sanity check for agents results. The floors differ across corpora because the task sets differ. Two tasks of the smaller corpus, one for root-cause analysis and one for customer triage, are declared abstentions: the corpus contains no cause.

\subsection{The Agent as Measurement Instrument}
\label{sec:method-probe}

We evaluate an \gls{llm}-driven agent as a measurement instrument. It operates in a ReAct-style loop~\cite{yao2023react}, alternating reasoning steps with the tool calls it has available through MCP: \texttt{tmf\_get} over the \gls{tmf} \glspl{api}, or \texttt{sparql\_query} over SPARQL on the graph. Because the agent's performance is a proxy for what each data representation offers, it carries no task-specific logic and no hand-coded traversal instructions. A \texttt{conclude} tool lets the agent state its final answer.

The prompt has two parts. The task guidance is the same for both interfaces: how to approach a question, how to treat an empty result, and how to report the answer. The interface part differs by interface: the routes and query conventions of the \gls{tmf} REST APIs, or the naming conventions of the graph and how it represents joins (Appendix~\ref{app:details}). Neither part gives counts, domain hints, or joins.

For every run we record the answer, each tool call and the tokens billed, and derive five measures. Two concern the answer. \emph{Accuracy} is the share of tasks solved, scored deterministically on the objects the agent names in \texttt{conclude}, which must include every object the declared answer requires, or any one where it accepts several (Appendix~\ref{app:details}). \emph{Precision} is the share of the named objects that are correct, which exposes answers that list candidates; we report it because the agent is a retrieval instrument and not a diagnostic specialist. Three concern the process. \emph{Data calls} is the number of tool calls a task takes, \emph{cost} the tokens per solved task, and \emph{evidence} whether every entity in the declared answer appeared in some tool result.

\subsection{Experiment Design}
\label{sec:method-arms}

The first factor of experiments is the \emph{interface}: the \gls{tmf} \glspl{api} via \gls{rest} as published, or the semantic layer, the lifted knowledge graph, queried with SPARQL, to assess what the semantic layer, taken as a whole, is worth to an agent. The graph is materialized, with the lifted data and the authored joins computed and stored before the agent queries it, and centralized, so the gains reported are an upper bound on performance for decentralized deployments, left for future work.

The second factor is the \emph{model}, the stronger gpt-5.6-sol or the weaker gpt-4.1 under the same prompt, to assess whether the benefit of the semantic layer depends on the capability of the model. The third is the \emph{corpus size}, a smaller and a larger corpus, to assess how scalability affects the conclusions.

The fourth is the \emph{task type}, root-cause analysis, customer impact or customer triage (Section~\ref{sec:method-answers}), to assess which questions depend on the joins; every cell runs all three. Crossing interface, model and corpus gives eight cells, and each cell runs every task twice. Table~\ref{tab:task-mix} gives the size of the two corpora and their tasks (see also Appendix~\ref{app:details}).

\begin{table}[h]
\centering
\caption{Size of the two corpora and their tasks: the same number of tasks, 72, in each, over networks of very different size.}
\label{tab:task-mix}
\footnotesize
\begin{tabular}{@{}lrr@{}}
\toprule
 & smaller & larger \\
\midrule
\multicolumn{3}{@{}l}{\emph{Network}} \\
Radio cells & \genSmallerCells & \genLargerCells \\
Generated data, MB & \corpusSmallerMegabytes & \corpusLargerMegabytes \\
Files served to the agent & \corpusSmallerServedfiles & \corpusLargerServedfiles \\
Triples in the lifted graph & \corpusSmallerTriples & \corpusLargerTriples \\
Resources & \corpusSmallerResources & \corpusLargerResources \\
Alarms & \corpusSmallerAlarms & \corpusLargerAlarms \\
Customers / services & \corpusSmallerCustomers\,/\,\corpusSmallerServices & \corpusLargerCustomers\,/\,\corpusLargerServices \\
\midrule
\multicolumn{3}{@{}l}{\emph{Joins}} \\
Declared by the generator & \joinsDeclared & \joinsDeclared \\
\quad of which unstated & \joinsTotal & \joinsTotal \\
\midrule
\multicolumn{3}{@{}l}{\emph{Tasks}} \\
Root-cause analysis & 24 & 24 \\
\quad of which accept any of several objects & \tasksSmallerAnyOf & \tasksLargerAnyOf \\
\quad of which abstention & \tasksSmallerAbstain & \tasksLargerAbstain \\
Customer impact & 24 & 24 \\
Customer triage & 24 & 24 \\
\quad of which concurrent causes & \tasksSmallerMulti & \tasksLargerMulti \\
Total & \tasksSmaller & \tasksLarger \\
\bottomrule
\end{tabular}
\end{table}

The larger corpus holds \corpusLargerTriples\ triples against \corpusSmallerTriples, for the same \tasksSmaller\ tasks, so what changes between corpora is the scale of data an agent must search and the answers it must find, not the type of questions it is asked.

\section{Results}
\label{sec:results}

\begin{table*}[t]
\centering
\caption{Tasks solved by interface and model on the smaller corpus (72 tasks) and a larger one with about \scaleFilesRatio\ times the served files (72 single-answer tasks), 2 trials each. \emph{Floor} is the best score of a trivial baseline. Cells: solved/attempted (\%). $\Delta$ is graph minus REST in percentage points, computed before rounding (green: graph better, red: worse). Last four rows: data-bearing tool calls per task, then thousands of billable tokens (uncached input plus output) per solved task. $^\dagger$: mean precision of the claimed objects below 50\% (the claim lists candidates; Table~\ref{tab:interface-robust}).}
\label{tab:interface-model}
\footnotesize
\begin{tabular*}{\textwidth}{@{\extracolsep{\fill}}lrrrrrrr@{}}
\toprule
 & & \multicolumn{3}{c}{gpt-5.6-sol} & \multicolumn{3}{c}{gpt-4.1} \\
\cmidrule(lr){3-5}\cmidrule(lr){6-8}
\textbf{Task type} & \textbf{Floor (\%)} & REST & Graph & $\Delta$ & REST & Graph & $\Delta$ \\
\midrule
\multicolumn{8}{l}{\emph{Smaller corpus}} \\
Customer impact (24) & 0 & 48/48 (100) & 48/48 (100) & 0 & 0/48 (0) & 1/48 (2) & \textcolor[HTML]{00B050}{\textbf{+2}} \\
Root-cause analysis (24) & 54 & 38/48 (79) & 40/48 (83) & \textcolor[HTML]{00B050}{\textbf{+4}} & 14/48 (29) & 37/48 (77) & \textcolor[HTML]{00B050}{\textbf{+48}} \\
Customer triage (24) & 28 & 38/48 (79) & 38/48 (79) & 0 & 23/48 (48)$^\dagger$ & 19/48 (40)$^\dagger$ & \textcolor[HTML]{D70000}{\textbf{$-$8}} \\
\midrule
All tasks & -- & 124/144 (86) & 126/144 (88) & \textcolor[HTML]{00B050}{\textbf{+1}} & 37/144 (26) & 57/144 (40) & \textcolor[HTML]{00B050}{\textbf{+14}} \\
\midrule
\multicolumn{8}{l}{\emph{Larger corpus}} \\
Customer impact (24) & 0 & 25/48 (52) & 48/48 (100) & \textcolor[HTML]{00B050}{\textbf{+48}} & 0/48 (0) & 0/48 (0) & 0 \\
Root-cause analysis (24) & 51 & 40/48 (83) & 48/48 (100) & \textcolor[HTML]{00B050}{\textbf{+17}} & 11/48 (23) & 32/48 (67) & \textcolor[HTML]{00B050}{\textbf{+44}} \\
Customer triage (24) & 5 & 48/48 (100) & 48/48 (100) & 0 & 15/48 (31)$^\dagger$ & 11/48 (23)$^\dagger$ & \textcolor[HTML]{D70000}{\textbf{$-$8}} \\
\midrule
All tasks & -- & 113/144 (78) & 144/144 (100) & \textcolor[HTML]{00B050}{\textbf{+22}} & 26/144 (18) & 43/144 (30) & \textcolor[HTML]{00B050}{\textbf{+12}} \\
\midrule
Data calls per task, smaller &  & 18.7 & 8.5 & \textcolor[HTML]{00B050}{\textbf{$-$10.2}} & 10.4 & 9.8 & \textcolor[HTML]{00B050}{\textbf{$-$0.6}} \\
Data calls per task, larger &  & 17.6 & 9.7 & \textcolor[HTML]{00B050}{\textbf{$-$7.9}} & 8.9 & 8.3 & \textcolor[HTML]{00B050}{\textbf{$-$0.6}} \\
Tokens per solved task (k), smaller &  & 28 & 36 & \textcolor[HTML]{D70000}{\textbf{+8}} & 29 & 29 & +0 \\
Tokens per solved task (k), larger &  & 39 & 35 & \textcolor[HTML]{00B050}{\textbf{$-$4}} & 61 & 44 & \textcolor[HTML]{00B050}{\textbf{$-$17}} \\
\bottomrule
\end{tabular*}
\end{table*}

The results answer two questions. The first is how far the published schemas take a consumer toward semantic network management (Section~\ref{sec:results-joins}). The second is what a standards-derived semantic layer over the telecom interfaces, built with authored rules, is worth to an AI agent, compared with the published \glspl{api}, and whether that depends on the model, the size of the data and the kind of question (Section~\ref{sec:results-interface}).

\subsection{Derivability of Cross-Specification Joins}
\label{sec:results-joins}

Section~\ref{sec:snm} argued that published schemas provide stable identity and traversable references (Levels~1 and~2), but not the references across specifications that Level~3 needs. Here we evaluate whether Level~3 can be derived from the schemas alone, measured against the generator's joins ground-truth (Table~\ref{tab:task-mix}).

Each join is a path of hops between its two fields. It is derivable only if a schema declares a link for every hop, so that a query can traverse the whole path from one field to the other. We classify each join by the first hop with no declared link. If the lifted ontology has no property for a field, we search the source schema text for that field name and count the omission against our extraction if it is present there, so the extraction-failure count is an upper bound.

This procedure measured that none of the \joinsTotal\ unstated joins, as identified by the data generator, can be derived from the published schemas: only \joinsHopsDeclared\ of the \joinsHopsTotal\ hops they require are declared by any schema. The schemas reach the identifier but not the link across specifications, and in every case the reason is one of three: the schema omits the field that carries the identifier, the surface publishes no schema at all, or the correspondence lives in prose or in a derivation that no schema could state (see Appendix~\ref{app:details}).

\subsection{Effect of the Semantic Layer on AI Agents}
\label{sec:results-interface}

\begin{figure*}[t]
\centering
\includegraphics[width=\textwidth]{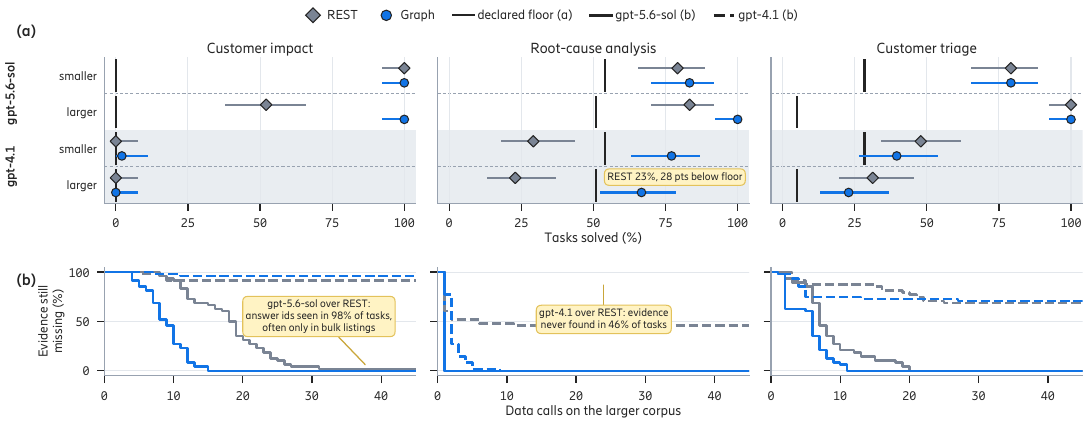}
\caption{Results by task type. (a) Tasks solved on the smaller and the larger corpus, for each model (shaded band and left label). Grey diamonds are \gls{rest} and blue circles the graph. Whiskers are 95\% Wilson intervals over the 48 attempts of a cell, which treat the two trials of a task as independent and are therefore optimistic. The black tick is the declared floor of that task type on that corpus. (b) Share of tasks whose answer evidence is still missing after $k$ data calls on the larger corpus. The axis runs past the longest retrieval (\traceLongestRetrieval\ calls), so a curve ending above zero marks tasks whose evidence was never retrieved.}
\label{fig:scale}
\end{figure*}

Table~\ref{tab:interface-model} gives the tasks solved for every combination of experiment factors. Table~\ref{tab:interface-margins} aggregates those numbers one factor at a time to get an estimate of each factor effect. Figure~\ref{fig:scale} shows the solved shares of each cell with their uncertainty (a) and where the evidence goes missing (b).

\begin{table}[!h]
\centering
\caption{Tasks solved (\%) by interface, pooled over the other three factors; $\Delta$ is graph minus REST in points.}
\label{tab:interface-margins}
\footnotesize
\begin{tabular}{@{}llrrrr@{}}
\toprule
 & & $n$ & REST & Graph & $\Delta$ \\
\midrule
\multirow{1}{*}{All} & every attempt & 576 & 52 & 64 & \textcolor[HTML]{00B050}{\textbf{+12}} \\
\midrule
\multirow{2}{*}{Model} & gpt-5.6-sol & 288 & 82 & 94 & \textcolor[HTML]{00B050}{\textbf{+11}} \\
 & gpt-4.1 & 288 & 22 & 35 & \textcolor[HTML]{00B050}{\textbf{+13}} \\
\midrule
\multirow{2}{*}{Corpus} & smaller & 288 & 56 & 64 & \textcolor[HTML]{00B050}{\textbf{+8}} \\
 & larger & 288 & 48 & 65 & \textcolor[HTML]{00B050}{\textbf{+17}} \\
\midrule
\multirow{3}{*}{Task type} & Customer impact & 192 & 38 & 51 & \textcolor[HTML]{00B050}{\textbf{+12}} \\
 & Root-cause analysis & 192 & 54 & 82 & \textcolor[HTML]{00B050}{\textbf{+28}} \\
 & Customer triage & 192 & 65 & 60 & \textcolor[HTML]{D70000}{\textbf{$-$4}} \\
\bottomrule
\end{tabular}
\end{table}

\paragraph{Interface} Overall, the semantic layer solves more tasks than the published \glspl{api} (Table~\ref{tab:interface-margins}). In the cells (Table~\ref{tab:interface-model}), the stronger model reaches its answers in about half as many data calls as over \gls{rest} at both corpus sizes (\strongGraphCalls\ against \strongRestCalls\ per task on the smaller corpus), but each call returns more data, so a task moves more data in total, not less (\strongGraphResultK k against \strongRestResultK k tokens of tool results per task). Whether this pays off depends on scale: on the smaller corpus a solved task is cheaper over \gls{rest} (\tokStrongRestXs k against \tokStrongGraphXs k), but as the corpus grows the cost over \gls{rest} rises (\tokStrongRestLarge k) and over the graph it does not (\tokStrongGraphLarge k).

\paragraph{Model} The semantic layer helps the two models by a similar amount (Table~\ref{tab:interface-margins}). In the cells (Table~\ref{tab:interface-model} and Figure~\ref{fig:scale}(a)), the help sits in different places: over the graph the weaker model solves more root-cause tasks than over \gls{rest} (\pctWeakRestXsRca\% to \pctWeakGraphXsRca\% on the smaller corpus, \pctWeakRestLargeRca\% to \pctWeakGraphLargeRca\% on the larger), and it solves almost no customer-impact task over either interface. Its cost per solved task also diverges with the data more sharply than the stronger model's (from \tokWeakRestXs k to \tokWeakRestLarge k over \gls{rest}, against \tokWeakGraphXs k to \tokWeakGraphLarge k over the graph).

\paragraph{Corpus size} The gain grows with the corpus (Table~\ref{tab:interface-margins}), and the stronger model carries the growth. In the cells (Table~\ref{tab:interface-model}), the interface hardly matters to it on the smaller corpus (\resStrongRestPct\% over \gls{rest}, \resStrongGraphPct\% over the graph), but as the corpus grows it comes to depend on the interface (Figure~\ref{fig:scale}(a)): over the graph it still solves every task (\pctStrongGraphLargeAll\%), while over \gls{rest} it falls to \pctStrongRestLargeAll\%.

\paragraph{Task type} The gain differs by task type (Table~\ref{tab:interface-margins}): it is largest for root-cause analysis, then customer impact, and absent for customer triage. In the cells (Table~\ref{tab:interface-model}), most of the stronger model's loss over \gls{rest} at scale is customer impact (\pctStrongRestLargeImpact\% over \gls{rest}). Customer triage does not separate the interfaces: the stronger model solves it equally often over both (\pctStrongRestXsTriage\% on the smaller corpus and \pctStrongRestLargeTriage\% on the larger), and the weaker model's triage scores, marked in Table~\ref{tab:interface-model}, overstate what it did: its claims list candidates, and on average at most \precFlaggedMax\% of the objects it names are causes (Table~\ref{tab:interface-robust}).

\subsection{Why the Agents Fail: Evidence from the Traces}
\label{sec:results-traces}

Figure~\ref{fig:scale}(b) shows, on the larger corpus, how many data calls a task takes before its evidence is complete. The traces behind it show where and why the agents fail.

\paragraph{Retrieval} The weaker model retrieves the evidence in only \traceEvidenceWeakRest\% of larger-corpus tasks over \gls{rest} and \traceEvidenceWeakGraph\% over the graph, against \traceEvidenceStrongRest\% and \traceEvidenceStrongGraph\% for the stronger model. Over \gls{rest} it filters collections by identifiers they do not hold, gets empty results and reads them as absence, so on most impact tasks it concludes that no customer is affected. Over the graph it finds the affected services but queries outward from them, and the customers are linked to a service only through the product that points to it.

\paragraph{Linkage} The stronger model retrieves the evidence over either interface, but in different forms. Over the graph the entities arrive linked to the task, and it never answers wrong once it holds them. Over \gls{rest}, on customer impact, they often arrive only inside listings of whole collections, without the link to the task, and \traceDecisionStrongRest\% of tasks end in a wrong answer despite the evidence. The failed runs share a pattern: every degraded cell carries two nationwide slice services besides its regional one, the agent finds one and misses the other, and with it the customer who uses only that service.

In both cases the failures sit at links that the published interfaces do not state: the weaker model does not find them, and the stronger model must rebuild them from listings over \gls{rest}, which the graph spares it.

\section{Conclusion}
\label{sec:conclusion}

Telecom standards already formalize much of the industry’s operational world, but not enough of its meaning. They define interfaces, payloads, and identifiers across business, service, and network domains, yet often stop short of declaring how those identifiers correspond across specifications. For AI agents expected to retrieve, connect, and reason across systems on their own, that omission becomes a structural limitation.

This paper examined that limitation through the lens of semantic network management. Using an end-to-end, standards-aligned 5G data view, we measured how far the published schemas go on their own, and what changes when the missing cross-domain semantics are made explicit in a semantic layer.

The result is clear: the schemas reach the identifiers and stop before the links. None of the 12 unstated joins in our evaluation can be derived from the published schemas alone. Every such correspondence must therefore be supplied either by authored domain rules or by LLM inference at run time.

That missing semantics is not just a modeling inconvenience; it is an operational cost. In our experiments, the semantic layer mattered most where agentic automation is hardest: larger corpora and questions that cross specifications. On the larger corpus, GPT-5.6-sol solved every task over the graph, but only 78\% over \gls{rest}, while using roughly half as many data calls. GPT-4.1 improved mainly on root-cause analysis and still failed on customer-impact tasks over both interfaces. As scale grows, what the standards leave unsaid increasingly determines what agents can retrieve, connect, and solve.

If these findings hold beyond a synthetic corpus, they argue for more than better middleware. They argue for standards that publish semantics as well as syntax: stable identifiers, typed references, and explicit cross-specification correspondences. Telecom may not need to reinvent its management stack for agentic AI, but it does need to finish specifying it. %

Two questions remain: whether the same gains hold on operational, multi-vendor data, and how much of the benefit comes from explicit semantics rather than query language alone. Answering them would turn semantic network management from a promising architecture into a concrete target for telecom standardization.

\bibliographystyle{IEEEtran}
\bibliography{references}

\appendices

\makeatletter
\setcounter{topnumber}{3}\renewcommand{\topfraction}{0.95}\renewcommand{\textfraction}{0.05}
\renewcommand{\floatpagefraction}{0.8}
\makeatother

\newcommand{\tabstyle}{\let\footnotesize\scriptsize}

\section{Experimental Details}
\label{app:details}

\subsection{Agent configuration}

Both arms run the same agent loop, with at most 25 turns and 200{,}000 billable tokens per question, and every tool the agent calls is an MCP tool. The REST arm is offered one data tool, \texttt{tmf\_get}, a GET over the \gls{tmf} routes with their query string, and the graph arm one, \texttt{sparql\_query}. Both are offered \texttt{conclude}, which records the answer as a list of entities with the cause and the evidence; an agent that stops without it is reminded once, which no run needed. Neither arm has memory or schema tools. The prompt has two parts. The task guidance is one hand-written text, identical in both prompts and for both models. The REST interface part lists the routes and their declared query parameters, read from the server's OpenAPI document, and the access conventions of the TM Forum REST design guidelines (TMF630): dereference by id, equality filtering on top-level attributes, projection and pagination. The graph interface part lists the classes the graph instantiates, where its ontology lives, its IRI and label conventions, and the merged nodes the join rules produce, which carry the types and identifiers of both systems. The graph also holds the agreement and account records of the corpus, which the REST server does not serve; the agents queried them but never used them in an answer, since they link only to product offerings. Both models are reached through the same gateway: gpt-5.6-sol is the
deployment name of the stronger model, run at its default settings because it accepts no temperature, and gpt-4.1 of the weaker, run at
temperature~0.

\begin{table}[t]
\begingroup\makeatletter
\setlength{\textwidth}{\columnwidth}\setlength{\tabcolsep}{1.6pt}\tabstyle
\renewenvironment{table}[1][]{\def\@captype{table}\centering\scriptsize}{}

\begin{table}[t]
\centering
\caption{Checks on Table~\ref{tab:interface-model}. \emph{Trials}: solved/attempted for each trial of each cell. \emph{Tokens per attempted task}: billable tokens (uncached input plus output) averaged over all rows, thousands. \emph{Mean precision}: per run, objects named that are in the answer over all objects named of the kind asked (objects directly above or below a correct one, such as a cell of the faulty DU, count as neither), averaged over the runs that named one; -- where fewer than 10 did.}
\label{tab:interface-robust}
\footnotesize
\begin{tabular*}{\textwidth}{@{\extracolsep{\fill}}lrrrr@{}}
\toprule
 & \multicolumn{2}{c}{gpt-5.6-sol} & \multicolumn{2}{c}{gpt-4.1} \\
\cmidrule(lr){2-3}\cmidrule(lr){4-5}
 & REST & Graph & REST & Graph \\
\midrule
\multicolumn{5}{l}{\emph{Trials (solved/attempted)}} \\
Smaller corpus, trial 0 & 64/72 & 63/72 & 20/72 & 28/72 \\
Smaller corpus, trial 1 & 60/72 & 63/72 & 17/72 & 29/72 \\
Larger corpus, trial 0 & 54/72 & 72/72 & 13/72 & 18/72 \\
Larger corpus, trial 1 & 59/72 & 72/72 & 13/72 & 25/72 \\
\midrule
\multicolumn{5}{l}{\emph{Tokens per attempted task (k)}} \\
Smaller corpus & 24 & 31 & 7 & 11 \\
Larger corpus & 31 & 35 & 11 & 13 \\
\midrule
\multicolumn{5}{l}{\emph{Mean precision of the claimed objects (\%)}} \\
Smaller, impact & 100 & 100 & -- & 100 \\
Smaller, root cause & 84 & 85 & 61 & 84 \\
Smaller, triage & 83 & 86 & 20 & 20 \\
Larger, impact & 100 & 100 & -- & -- \\
Larger, root cause & 100 & 100 & 92 & 100 \\
Larger, triage & 100 & 98 & 9 & 5 \\
\bottomrule
\end{tabular*}
\end{table}

\makeatother\endgroup
\end{table}

\subsection{Scoring}

A run is scored on the entities it passes to \texttt{conclude}, not on its prose. Each claimed identifier is mapped to the record it names through an alias table built from the corpus itself: every inventory record carries its id, its name and its distinguished name, the identifiers the agents see, and an instance IRI is reduced to its last segment. A DU's managed-element name is accepted for the DU, because each managed element hosts exactly one DU. A task is solved when the claim includes every object its answer requires; where the answer accepts any of several objects, or lists the causes of concurrent incidents, any one suffices. The abstention task is solved only by an empty claim. \emph{Precision} is the share of claimed objects of the asked kind (managed objects, or customers and services for impact) that are in the answer; an object directly above or below a correct one, such as a cell of the faulty DU, counts as neither right nor wrong. Table~\ref{tab:interface-robust} gives the mean precision of each cell: it falls below \precFlagPct\% only for the weaker model's triage, whose claims list candidates (at most \precFlaggedMax\%), and is at least \precOthersMin\% elsewhere.

\subsection{Example tasks}

Table~\ref{tab:example-tasks} gives one task of each type from the smaller corpus, as the agent receives it, with the answer the generator declares. All three are built on the same incident, which raised the four alarms; triage starts from a customer instead of the alarms, and its window also holds the alarms of two other incidents.

\begin{table}[t]
\caption{One task per type, smaller corpus (each question is preceded by its task id in the released corpus; the timestamp is shortened).}
\label{tab:example-tasks}
\begingroup\tabstyle\scriptsize\setlength{\tabcolsep}{2pt}
\begin{tabular}{@{}p{0.13\columnwidth}p{0.42\columnwidth}p{0.42\columnwidth}@{}}
\hline
Type & Question & Declared answer \\
\hline
Root cause &
\texttt{RCA-\allowbreak INC-\allowbreak RegionalSubnet-\allowbreak A1-1}. ``4 cell alarms were raised in RegionalSubnet-A1 within 300\,s. What caused them, and on which managed object?'' &
Three managed objects of one gNB-DU, \texttt{NR01gNodeBRadio00001}: its F1-C and F1-U connections and its DU function. Two other incidents occur elsewhere in the same window. \\
Customer impact &
\texttt{IMPACT-\allowbreak RegionalSubnet-\allowbreak A1-1}. ``4 alarms were raised between 09:26:02Z and 09:26:02Z. Which customers and which services are affected?'' &
Services \texttt{SVC-NATIONWIDE-00001} and \texttt{SVC-REGIONAL-00001}; customers \texttt{CUST-00001} and \texttt{CUST-00002}, 2 of the 5 customers in the corpus. The path runs from each cell to the services that rely on it, then to the customers whose products realize them. \\
Customer triage &
\texttt{TRIAGE-\allowbreak RegionalSubnet-\allowbreak A1-1-\allowbreak CUST-00002}. ``Customer CUST-00002 reports degraded service between 09:26:02Z and 09:26:02Z. What caused it, and on which managed object?'' &
The same three managed objects as the root-cause task. The window holds three incidents, so a random guess is right a third of the time; the customer is what tells them apart. \\
\hline
\end{tabular}
\endgroup
\end{table}

\subsection{Corpus sizes and the joins}

Table~\ref{tab:corpus-sizes} gives the absolute size of each corpus. The trees served to the agents keep a subset of the
generated files: \corpusSmallerServedfiles\ of \genSmallerFiles\ for the smaller corpus and \corpusLargerServedfiles\ of \genLargerFiles\ for the larger. The identity grade is built from three kinds of identifier field (a list
of mapping identifiers, an external identifier, and an \texttt{id} that is itself a distinguished name). Table~\ref{tab:join-registry}
groups the \joinsTotal\ unstated joins by why none is derivable. The evaluated
graph uses three ontologies of \tboxTriples, \ctsTboxTriples\ and \nrmTboxTriples\ triples, from the standard, the vendor schema and the 3GPP
resource model; \overlayTriples\ of the first are the manually curated overlay.

\begingroup\tabstyle\renewcommand{\textbf}[1]{#1}\begin{table}[t]
\centering
\caption{Why the twelve declared joins that no payload field states cannot be derived from the published schemas.
\emph{Hops} counts the hops a schema declares out of the hops the joins in the row traverse. The per-join detail is in
the experiment output.}
\label{tab:join-registry}
\footnotesize
\begin{tabular}{@{}>{\raggedright\arraybackslash}p{0.25\columnwidth}cc
>{\raggedright\arraybackslash}p{0.47\columnwidth}@{}}
\toprule
\textbf{Why not derivable} & \textbf{Joins} & \textbf{Hops} & \textbf{What the schemas leave unsaid} \\
\midrule
The join key is absent from the schema & 4 & 7/17 & Where a topology object's distinguished name travels: in a free-form attribute map, or inside the \texttt{id} string of a TMF resource. No schema describes either. \\
No schema is published & 4 & 0/8 & The vendor's assurance definitions name cells, network functions, slice subnets and tracking areas; no machine-readable schema says which topology objects those names denote. \\
The rule must be authored & 4 & 0/9 & Four derivations: a KPI's display name from its identifier suffix, the alias form of a cell's name, the topology attribute a KPI is bound to, and a KPI's slice scope. \\
\midrule
All & 12 & 7/34 & \\
\bottomrule
\end{tabular}
\end{table}
\endgroup

\subsection{Size categories}

The generator ships presets that form a ladder by the number of cells. We evaluate two sizes only for convenience, starred in
Table~\ref{tab:corpus-sizes}: a smaller corpus (the XS preset with four TACs and six gNB-DUs per TAC, 96 cells) and a larger one (the S preset
with 20 regions, \genLargerCells\ cells). The other columns are built with the same dials: S with its default five regions, and M, which has
\genMCells\ cells, \genMRatio\ times the larger corpus, and generates in \genMSeconds\ seconds on one machine. The two smallest presets
cannot carry the task dials, and the TAC and DU counts are open knobs beyond the presets.

\begingroup\tabstyle\begin{table}[!t]
\centering
\caption{Size of the generated corpora, built with the campaign's dials. The two starred, bold columns are the corpora evaluated here, chosen for convenience; S and M show what the same generator produces at larger scale. Cells are declared by the preset; files, megabytes, seconds and entity counts are measured on one machine from the generated tree. The trees served to the agents keep a subset of these files (see text).}
\label{tab:corpus-sizes}
\footnotesize
\begin{tabular}{@{}lrrrr@{}}
\toprule
 & $\star$ \textbf{smaller} & S & $\star$ \textbf{larger} & M \\
\midrule
Cells & \textbf{96} & 2{,}720 & \textbf{2{,}720} & 7{,}200 \\
Files & \textbf{826} & 8{,}301 & \textbf{11{,}124} & 21{,}885 \\
MB & \textbf{8} & 217 & \textbf{321} & 711 \\
Seconds & \textbf{9.3} & 372 & \textbf{448} & 1{,}854 \\
Customers & \textbf{5} & 6 & \textbf{21} & 6 \\
Services & \textbf{5} & 7 & \textbf{22} & 7 \\
Products & \textbf{90} & 108 & \textbf{378} & 108 \\
Resources & \textbf{180} & 5{,}280 & \textbf{6{,}480} & 15{,}200 \\
Alarms & \textbf{158} & 2{,}283 & \textbf{3{,}006} & 5{,}867 \\
\bottomrule
\end{tabular}
\end{table}
\endgroup

\section{Additional Results}
\label{app:results}

Table~\ref{tab:interface-robust} repeats the results with the checks a reader needs. The two trials of a cell differ by at most \trialMaxDiff\ of its 72 tasks. The one abstention task, whose three cell alarms have no cause in the data, was never answered correctly: in all eight runs, two per condition, the agent named the cells or their DU. The graph's entry point is the identifier inside the IRI. Probes run without the agent (Table~\ref{tab:trace-probes}, by entity type) find an entity by identifier as a \gls{rest} filter in \probeRestIdOk\ of \probeRestIdN\ tries, and through the IRI built from it in \probeGraphIriOk\ of \probeGraphIriN, but never through a label (\probeGraphLabelOk\ of \probeGraphLabelN), because labels hold names. The graph prompt states this convention; without it, an agent holding only an identifier would have no way in.

\begingroup\tabstyle\begin{table}[t]
\centering
\caption{Naive-client battery: for 20 real entities of each type, the number found by each plausible first move, with no agent involved. A dash means the move does not apply: alarms carry no label or name, and the graph has no resource class.}
\label{tab:trace-probes}
\footnotesize
\begin{tabular}{@{}lrrrrr@{}}
\toprule
\textbf{Move} & customer & service & alarm & resource & product \\
\midrule
\multicolumn{6}{l}{\emph{REST}} \\
\texttt{?id=\textless{}id\textgreater{}} & 20/20 & 20/20 & 20/20 & 20/20 & 20/20 \\
\texttt{?name=\textless{}id\textgreater{}} & 0/20 & 0/20 & 0/20 & 0/20 & 0/20 \\
\texttt{?name=\textless{}name\textgreater{}} & 20/20 & 20/20 & -- & 20/20 & 20/20 \\
\midrule
\multicolumn{6}{l}{\emph{Graph}} \\
\texttt{\textless{}IRI\textgreater{}} & 20/20 & 20/20 & 20/20 & -- & 20/20 \\
\texttt{rdfs:label "\textless{}id\textgreater{}"} & 0/20 & 0/20 & 0/20 & -- & 0/20 \\
\texttt{rdfs:label "\textless{}label\textgreater{}"} & 20/20 & 20/20 & -- & -- & 20/20 \\
\texttt{?s ?p "\textless{}id\textgreater{}"} & 0/20 & 0/20 & 0/20 & -- & 20/20 \\
\bottomrule
\end{tabular}
\end{table}
\endgroup

\emph{Entailment.} To ask whether generic inference would have helped, \gls{rdfs} entailment (type closure over subclasses, domain
and range typing, no \texttt{owl:sameAs} expansion) was applied to the contracted graph and the agents' own queries were replayed in
memory with and without the entailed triples. It adds \entailSmallerTriples\ triples to the smaller graph (\entailSmallerPct\%) and
\entailLargerTriples\ to the larger (\entailLargerPct\%), all \texttt{rdf:type}, filling classes that had no instances, such as every
resource and every cell. On the smaller corpus all \entailSmallerQueries\ replayable queries were replayed, and the replay matches the
recorded row counts on \entailSmallerAgree\ of the \entailSmallerCompared\ SELECT and ASK queries; \entailSmallerChanged\ change their result and \entailSmallerEmptyToNon\ go from empty to
non-empty. On the larger corpus a stratified sample of \entailLargerQueries\ replayable queries was replayed, \entailLargerChanged\ change
and \entailLargerEmptyToNon\ go from empty to non-empty. Queries the trace stored cut off cannot be replayed and are excluded.

\end{document}